# A One-Way Function Based On The Extended Euclidean Algorithm

Ephraim Feig, *Fellow IEEE* and Vivian Feig

**Abstract:** A problem based on the Extended Euclidean Algorithm applied to a class of polynomials with many factors is presented and believed to be hard. If so, it is a one-way function well suited for applications in digital signatures.

Let $P(x)$ and $Q(x)$ be polynomials defined over the Integers modulo a prime integer $p$ with $P(x)Q(x) = (x^{p-1} - 1) \mod p$. Also let $A(x)$ be a polynomial such that $A(x)P(x) \equiv 1 \mod Q(x) \mod p$. In general, the pair $\{P(x), Q(x)\}$ is not uniquely determined (mod $p$) from just $A(x)$ and $p$; using Mathematica, we have found counterexamples. However, when $\deg P(x) = \deg Q(x) = (p-1)/2$, we have not found counterexamples, but neither have we proved that the solution is unique. More pertinent for this correspondence is the question: is finding such a pair of polynomials a hard problem? That is, can one find such a pair from just $A(x)$ and $p$ without testing all possible factorizations of $x^{p-1} - 1$ and checking for the congruence until a match is found? Because

$$x^{p-1} - 1 = \prod_{j=1}^{p-1}(x - j) \mod p, \qquad (1)$$

the number of such pairs is exponential in $p$. Identity (1) is true because, by Fermat's Little Theorem [1], for every integer $j$ relatively prime to $p$, $j^{p-1} \equiv 1 \mod p$, and so each of the $p-1$ values of $j$ between 1 and $p-1$ is a root of the equation $x^{p-1} - 1 = 0 \mod p$.

If the problem is indeed hard, a one-way function [2] can be obtained as follows. Randomly factor $x^{p-1} - 1$ into two polynomials of degree $(p-1)/2$, and then use the Extended Euclidean Algorithm (EEA) [3] to find polynomials $A(x)$ and $B(x)$ such that

$$A(x)P(x) + B(x)Q(x) = 1 \mod p. \qquad (2)$$

From equation (2) we get $A(x)P(x) \equiv 1 \mod Q(x) \mod p$. Since computing the EEA is simple (polynomial in the size of the input), the problem of computing $A(x)$ from $\{P(x), Q(x)\}$ is simple; but its inverse, the computation of $\{P(x), Q(x)\}$ from $A(x)$ is assumed hard. The application of one-way functions to digital signatures is straightforward [4]. A would-be signer publishes $A(x)$ as a public key. To prove authenticity, the signer presents $P(x)$. Since, as assumed, the problem of determining $P(x)$ from $A(x)$ is hard, we accept the signer as authentic.

**References:**

1. T. Nagell, *Introduction to Number Theory*, Wiley, pp. 71-73, 1951.
2. O. Goldreich, *Foundations of Cryptography: Vol. 1, Basic Tools*. Cambridge University Press, 2001.
3. D. Knuth, *The Art of Computer Programming, Vol. 2: Seminumerical Algorithms,* Addison-Wesley, Reading, Mass., 3$^{rd}$ Edition, 1998.
4. L. Lamport, "Constructing digital signatures from a one-way function," Technical Report CSL-98, SRI International, Oct. 1979.